\documentclass[10pt,twocolumn,aps,prl,balancelastpage,longbibliography]{revtex4-1}
\usepackage[latin9]{inputenc}
\usepackage{color}
\usepackage{verbatim}
\usepackage{amsmath}
\usepackage{amssymb,ulem}
\usepackage{graphicx,cellspace,booktabs}
\usepackage{esint}
\usepackage[unicode=true,pdfusetitle,
 bookmarks=true,bookmarksnumbered=false,bookmarksopen=false,
 breaklinks=false,pdfborder={0 0 1},backref=false,colorlinks=true]{hyperref}
\usepackage{breakurl}
\usepackage{epstopdf}
\usepackage{feynmp-auto}
\usepackage{youngtab}
\usepackage{ytableau}

\makeatletter

\@ifundefined{textcolor}{}
{%
 \definecolor{BLACK}{gray}{0}
 \definecolor{WHITE}{gray}{1}
 \definecolor{RED}{rgb}{1,0,0}
 \definecolor{GREEN}{rgb}{0,1,0}
 \definecolor{BLUE}{rgb}{0,0,1}
 \definecolor{CYAN}{cmyk}{1,0,0,0}
 \definecolor{MAGENTA}{cmyk}{0,1,0,0}
 \definecolor{YELLOW}{cmyk}{0,0,1,0}
}

\usepackage[caption=false]{subfig}
\usepackage{bm}

\newcommand{\ket}[1]{|#1\rangle}

\newcommand{\bv}[1]{\mathbf{#1}}

\renewcommand{\(}{\left(}
\renewcommand{\)}{\right)}

\def\l@subsubsection#1#2{}
\makeatother

\begin{document}

\title{Generalized $U(1)$ Gauge Field Theories and Fractal Dynamics}

\author{Daniel Bulmash}
\author{Maissam Barkeshli}

\affiliation{Condensed Matter Theory Center and Joint Quantum Institute, Department of Physics, University of Maryland, College Park, Maryland 20472 USA}

\date{\today}
\begin{abstract}
We present a theoretical framework for a class of generalized $U(1)$ gauge effective field theories. These theories are defined
by specifying geometric patterns of charge configurations that can be created by local operators, which then lead to a
class of generalized Gauss law constraints. The charge and magnetic excitations in these theories have restricted, 
subdimensional dynamics, providing a generalization of recently studied higher-rank symmetric $U(1)$ gauge 
theories to the case where arbitrary spatial rotational symmetries are broken. These theories can describe situations 
where charges exist at the corners of fractal operators, thus providing a continuum effective field theoretic description of 
Haah's code and Yoshida's Sierpinski prism model. We also present a $3+1$-dimensional $U(1)$ theory that does not have a non-trivial
discrete $\mathbb{Z}_p$ counterpart.
\end{abstract}
\maketitle

It has recently been discovered that phases of matter can exist in which the dynamics of topologically non-trivial excitations are confined to 
subdimensional manifolds in a variety of novel ways. This phenomenon has been demonstrated in two contexts: 
(1) a series \cite{ChamonGlass, BravyiChamonModel, HaahsCode, YoshidaFractal,VijayFractons,VijayGaugedSubsystem,HsiehFractonsPartons, SlagleDuality} of (3+1)D 
lattice models of gapped Hamiltonians where the emergent topological excitations have restricted motion and are referred to as ``fractons,''
and (2) higher-rank symmetric $U(1)$ gauge field theories \cite{XuFractons1, XuFractons2, RasmussenFractons, PretkoSubdimensional,BulmashHiggs}. 
These developments have led to much recent activity \cite{NandkishoreFractonReview, VijayLayer, MaLayer, RegnaultLayer, PretkoElectromagnetism, PretkoWitten, VijayGaugedSubsystem, Williamson2016, HalaszFractons, PretkoElasticity, GromovElasticity, SlagleGenericLattices, ShirleyXCubeFoliations,SlagleFieldTheory, EmergentPhasesFractons, ShiFractonEntanglement, FractonEntanglement, RecoverableInformation, FractonCorrFunctions, VijayNonAbelianFractons, SongTwisted, YouSymmetricFracton}.

In this paper, we develop a class of generalized $U(1)$ gauge theories, which can in particular describe situations 
where the dynamics of the charged excitations are associated with fractal operators; in such cases, the energy cost to
creating isolated charges is exponentially large in their separation \cite{HaahU1Code}. The Higgs phases \cite{BulmashHiggs,MaHiggs} of these 
theories provide an effective field theoretic description of models such as Haah's code and Yoshida's Sierpinski prism model,
which have so far have lacked a description in terms of any effective continuum gauge theory. We also demonstrate the possibility
of non-trivial models whose $\mathbb{Z}_p$ counterparts have fully mobile particles and are thus conventional phases. 

An essential observation is that the possible motion of charges is fully determined by the set of charge 
configurations created by local operators, defined at some cutoff scale. For example, an isolated charge 
is mobile in the $x$ direction if and only if an $x$-oriented dipole can be created by a local operator. 
This suggests a perspective whereby a gauge theory can be specified, at least partially, in terms of a set of allowed geometric patterns of
charge configurations that can be created by local operators at the cutoff scale. Given a set of $U(1)$ charge 
configurations, when can a sensible (continuum) effective gauge theory be constructed? In this paper, 
we discuss how to construct continuum Gauss' Laws given a desired set of cutoff-scale charge configurations,
and derive the resulting gauge transformations, magnetic fields, effective Hamiltonians, and conserved quantities. 
We also prove, for the cases of one or two charge flavors, under which conditions there exists a well-defined magnetic field, leading to a nondegenerate Maxwell-type gauge theory. 

\begin{figure}[tb]
\centering
\hfill
\subfloat[\label{fig:RightPyramidCharges}]{\includegraphics[width=2cm]{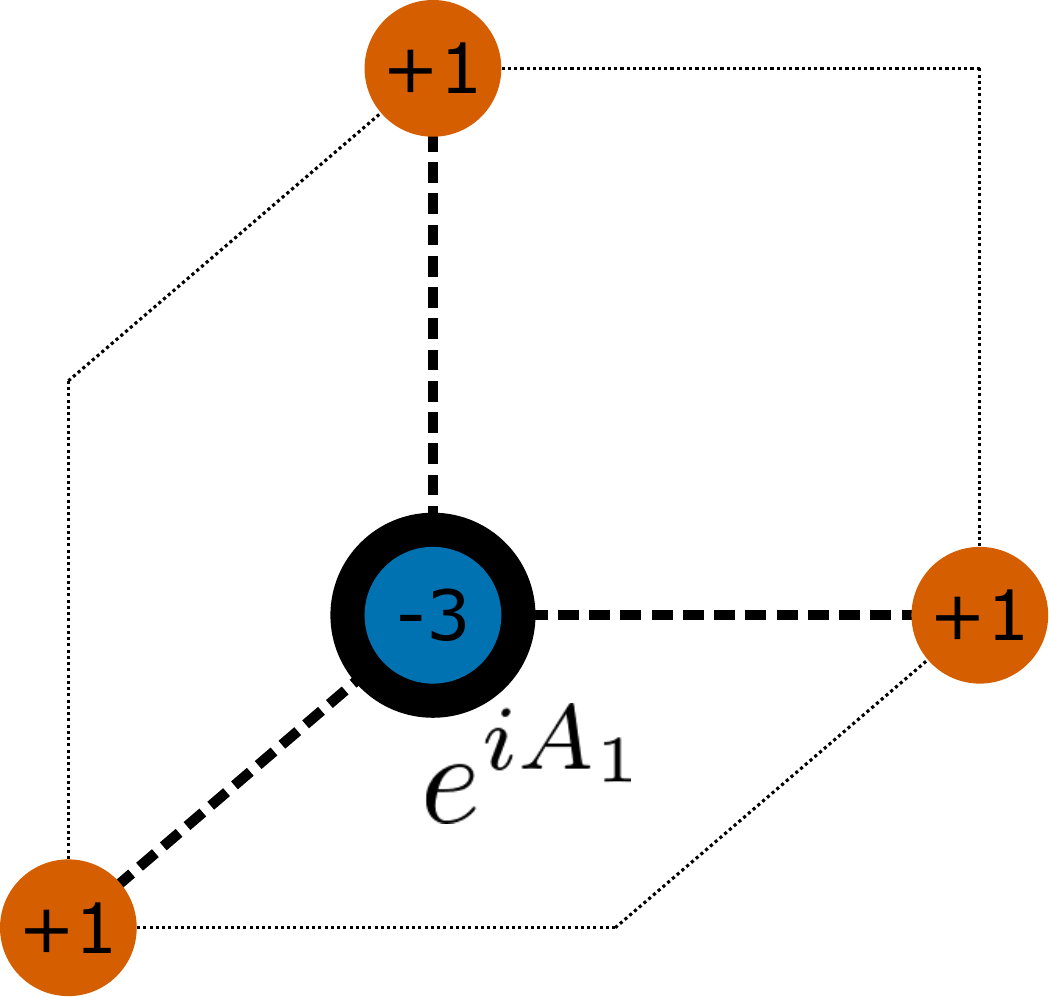}} \hfill
\subfloat[\label{fig:AngledPyramidCharges}]{\includegraphics[width=2cm]{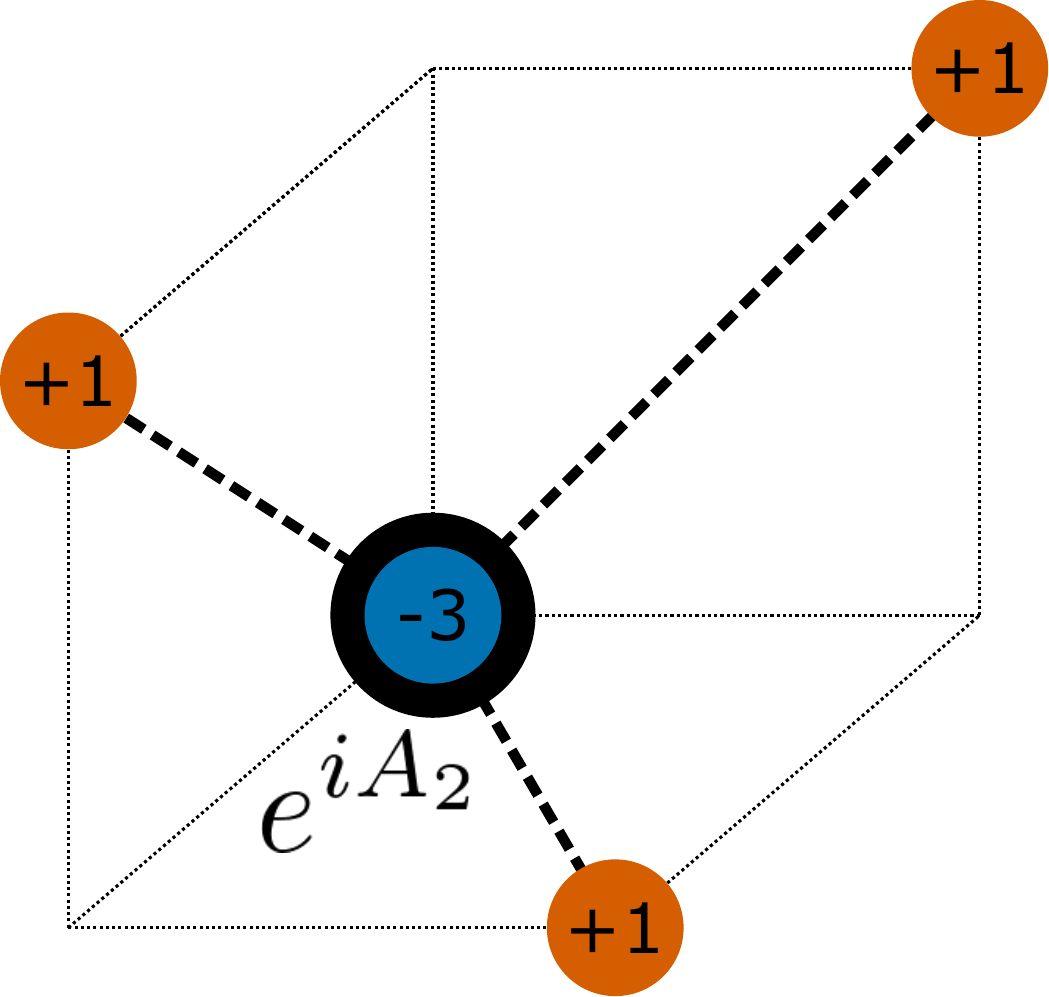}} \hfill
\subfloat[\label{fig:IsolatedHaahCharge}]{\includegraphics[width=4cm]{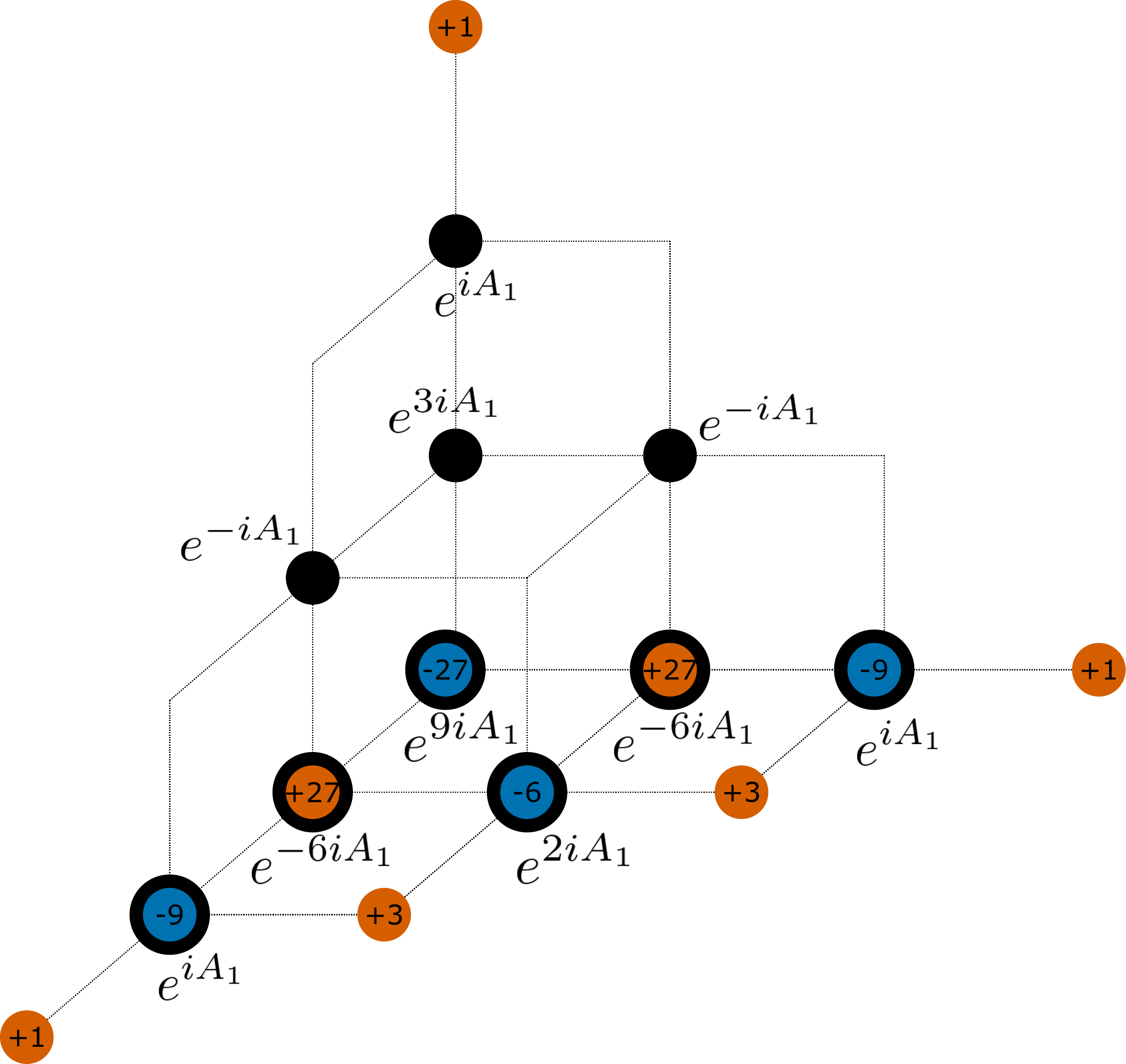}}
\caption{Charge configurations created in our $U(1)$ ``Haah's Code" gauge theory. Positive charges are orange, negative charges are blue, and black circles indicate the location of a local operator acts. (a) and (b): Generating charge configurations, created by the local action of $e^{iA_1}$ and $e^{iA_2}$ respectively. (c): Typical charge configuration with an isolated $+1$ charge created with repeated application of $e^{iA_1}$ (we have set $a_0=1$ for legibility). Isolating one charge requires a ``sheet" of charge of linear size $d$ to be created at distance $d$.}
\label{fig:HaahCharges}
\end{figure}

\textit{An Example - Haah's Code---} Let us consider a real scalar field $A_1$ with canonical conjugate $E_1$, i.e.
$[A_1(\bv{r}),E_1(\bv{r}'] = i\delta^3(\bv{r}-\bv{r}')$. For $A_1$ to be a gauge field, we consider the following
``Gauss Law'' constraint:
\begin{equation}
\rho(\bv{r}) = \sum_{i=1}^3 \partial_i E_1(\bv{r}),
\end{equation}
with $\rho$ the charge density. To understand the meaning of this Gauss Law, consider its
lattice regularization $\rho(\bv{r})/a_0 =  -3 E_1(\bv{r}) + \sum_{i=1}^3 E_1(\bv{r}+a_0 \hat{\bv{x}}_i)$, 
where $a_0$ is the lattice spacing. The action of the local operator $\exp(iA_1(\bv{r}) a_0)$, which acts as a raising operator for $E_1(\bv{r})$,
modifies all $\rho({\bf r}')$ that contain $E({\bf r})$ in its Gauss law constraint. The resulting charge configuration
is the tetrahedron of Fig. \ref{fig:RightPyramidCharges}. 

The Gauss Law constraint requires that
$\exp \left[\int d^3 \bv{r} \alpha(\bv{r})\left(\rho(\bv{r}) - \sum_{i=1}^3 \partial_i E_1(\bv{r})\right)\right] \ket{\psi} = \ket{\psi}$ for any state $\ket{\psi}$.
It is straightforward to check that this operator performs the ``gauge transformation" 
$A_1 \rightarrow A_1 + \sum_i \partial_i \alpha$, $\phi \rightarrow \phi - \alpha$, where $\phi$ is the phase field 
canonically conjugate to $\rho$. However, we see that there is no gauge-invariant operator consisting entirely of 
$A_1$ and its derivatives, so although there is an electric field $E_1$ in the theory, we cannot create a 
gauge-invariant magnetic field. Any gauge-invariant Hamiltonian would therefore be highly degenerate and unstable. 

The problem is that the number of independent gauge transformations, namely, one, is equal to the number of 
degrees of freedom. Thus let us conisder a second conjugate pair $A_2$ and $E_2$, such that at the lattice scale
$\exp(i A_2 a_0)$ creates a different charge configuration, which we choose to be the tetrahedron shown in Fig. 
\ref{fig:AngledPyramidCharges}. This can be achieved by modifying Gauss' Law to
\begin{equation}
\rho =  D_1 E_1 + D_2 E_2
\label{eqn:HaahGauss}
\end{equation}
with the differential operators $D_{1,2}$ defined by
\begin{align}
D_1 = \sum_i \partial_i, \;\;
D_2 = a_0\sum_{i<j}\partial_i \partial_j - 2 \sum_i \partial_i .
\end{align}
Upon discretizing, it is straightforward to check that $\exp(i A_2 a_0)$ and $\exp(i A_1 a_0)$ create the two tetrahedral configurations in Fig. \ref{fig:RightPyramidCharges}
and \ref{fig:AngledPyramidCharges}. We note this Gauss Law has $SO(2)$ rotational symmetry about the $(111)$ axis.

Now the gauge transformation law reads, for $l=1,2$,
$A_l \rightarrow A_l - \tilde{D}_l \alpha$,  $\phi \rightarrow \phi - \alpha$, where $\tilde{D}_l$ accounts for integration by parts:
$\tilde{D}_1 = -\sum_i \partial_i$, $\tilde{D}_2 = a_0\sum_{i<j}\partial_i \partial_j + 2 \sum_i \partial_i$.
There now exists a gauge-invariant magnetic field operator $B = \tilde{D}_1 A_2 - \tilde{D}_2 A_1$, and we can give this theory a Maxwell-like Hamiltonian
density 
\begin{align}
\mathcal{H}_{\text{G}} = \sum_i E_i^2 +  \frac{1}{2g^2}B^2,
\end{align}
where for simplicity we assume $E_1^2$ and $E_2^2$ have equal coefficients. 
The ``photon" has dispersion $\omega^2 = (a_0)^2\left(\sum_{i<j} k_ik_j\right)^2+5 \left(\sum_i k_i\right)^2$
which is gapped everywhere except at $k=0$ and has linear dispersion at small momentum except for soft 
quadratic dispersion along lines where $\sum_i k_i =0$.  

We can also couple the gauge theory covariantly to a gapped matter sector, for instance
a charge $p$ scalar matter field, with the total Hamiltonian $\mathcal{H} = \mathcal{H}_{\text{G}} + \mathcal{H}_{\text{M}}$, and 
\begin{align}
\mathcal{H}_{M} = \frac{L^2}{2M} - \sum_{i=1,2}V_i \cos(\tilde{D}_i\phi - p A_i) 
\end{align}
$\phi$ the phase field for a charge-$p$ bosonic matter field and $L$ is its conjugate 
number operator. This theory is subject to the constraint $D_1 E_1 + D_2 E_2 = p L$.
The corresponding Lagrangian follows straightforwardly.

In standard $U(1)$ gauge theory, there is one conserved quantity, the total charge. 
The total charge is also conserved in our Haah's code gauge field theory, but that is not the only conserved quantity 
- remarkably, there are infinitely many conserved quantities. Rotating to orthonormal coordinates $u,v,w$ with $u$ along 
the $(111)$ axis (the choice of $v$ and $w$ are arbitrary by $SO(2)$ rotational symmetry), it can be checked from Gauss' 
Law that in the absence of spatial boundaries
\begin{equation}
Q_f \equiv \int du dv dw f(v,w) \rho(u,v,w) = 0
\end{equation}
whenever $f$ is harmonic, that is, $(\partial_v^2 + \partial_w^2)f(v,w)=0$. 
Therefore, for all harmonic functions $f$, the quantity $Q_f$, which has density $f(v,w)\rho(u,v,w)$, is conserved.

\begin{figure}
	\includegraphics[width=8cm]{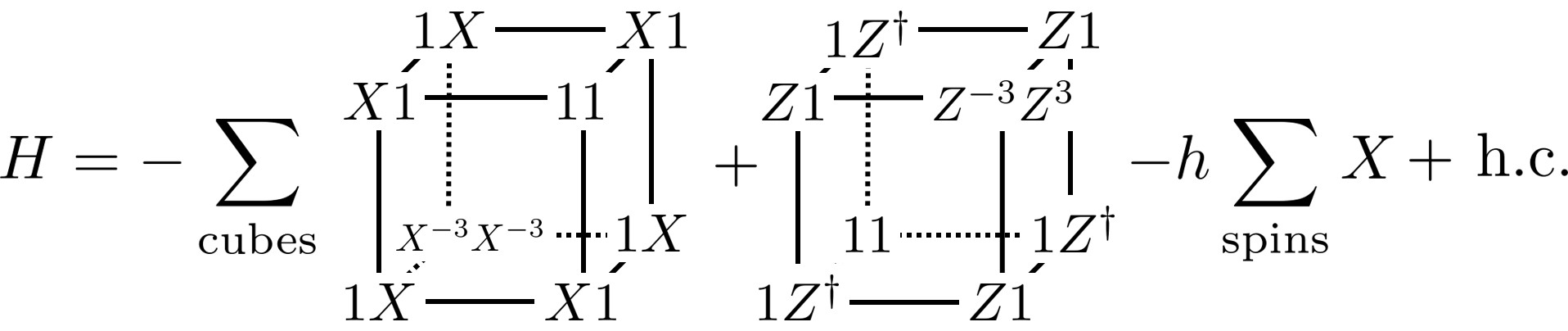}
	\caption{Hamiltonian for our chosen $\mathbb{Z}_p$ generalization of Haah's code, also used in Ref. \cite{HaahU1Code}. There is one of each term per elementary cube of the cubic lattice, two $N$-component spins per site, and each term is a product of five generalized Pauli operators. }
	\label{fig:HaahHamiltonian}
\end{figure}

If the theory is regularized on a cubic lattice using the discretization procedure we described earlier, we find that our model 
is the $U(1)$ generalization of Haah's code considered by Haah \cite{HaahU1Code}. That model can be obtained by taking 
the $p \rightarrow \infty$ limit of one natural $\mathbb{Z}_p$ generalization \footnote{There is not a canonical $U(1)$ generalization of the $\mathbb{Z}_2$ Haah's code, but the choice we consider is convenient due to the $SO(2)$ rotational symmetry about the $(111)$ axis.} of Haah's code with the Hamiltonian shown in Fig. \ref{fig:HaahHamiltonian}. 
As such, our construction yields an effective field theory for the $U(1)$ generalization of Haah's code. Furthermore, following 
the same procedure as in Refs. \cite{BulmashHiggs, MaHiggs}, condensing charge-$p$ objects by considering the limit $V_i \gg 1/2M$ 
in the $U(1)$ theory breaks the $U(1)$ gauge symmetry down to $\mathbb{Z}_p$, yielding the $\mathbb{Z}_p$ Haah's code. 
In this limit we can describe fluctuations of the phase field $\phi$ by expanding the cosine to quadratic order:
$\mathcal{H}_{\text{M}} \approx \frac{1}{2}\sum_i V_i (\tilde{D}_i\phi - p A_i)^2$. We thus propose $\mathcal{H} = \mathcal{H}_{\text{G}} + \mathcal{H}_{\text{M}}$
as an effective field theory for the $\mathbb{Z}_p$ Haah's code.

The above field theory can thus describe the gapless ``photon'' field, and perturbative fluctuations of the matter sector about the 
ground state of the Higgsed phase. Nevertheless, we see that the lattice regularization plays several crucial roles, in contrast to standard one-form
$U(1)$ gauge field theories. The lattice cutoff $a_0$ appears in Gauss' Law and cannot be removed by redefining variables. If we consider
$a_0\rightarrow 0$, in Gauss' Law, then $A_2 + 2A_1$ becomes gauge invariant and thus the remaining degrees of freedom 
in the theory are pure gauge. Hence, to maintain the correct gauge structure, it is crucial to think of this field theory as 
an \textit{effective} field theory with the lattice cutoff playing an important role. 

The lattice regularization is also required to express Wilson-type operators for the charges in terms of the fundamental fields. 
In this case, we see that attempting to isolate a charge on the lattice with repeated applications of $e^{iA_1}$ results in the 
charge configuration shown in Fig. \ref{fig:IsolatedHaahCharge}, which requires a divergent energy cost. Haah has shown \cite{HaahU1Code} 
that in the $U(1)$ lattice model, the energy to keep a charge a distance $d$ from all other charges goes as $\exp(d/a_0)$, 
and is thus infinite in the continuum limit. 

\textit{General Construction---}
To generalize the above construction, we consider two pieces of input data for a generalized $U(1)$ gauge theory: 
(1) A list of distinct types (flavors) of charges $\lbrace a \rbrace$, and (2) A generating set of charge configurations $\lbrace l \rbrace$.
By ``generating set of charge configurations" we mean a set of charge configurations, each with finitely many charges, 
such that they and their translates (we assume translation symmetry) are a basis under the operation of superposition 
for the space of all charge configurations in the theory.
To the $l$th generating charge configuration, we associate one component $A_l$ of the gauge field, 
along with its conjugate variable $E_l$ (``electric field"), $[A_l(\bv{r}),E_m(\bv{r}')] = i\delta_{lm}\delta^d(\bv{r}-\bv{r}')$.
The indices $l,m$ need not be related to spacetime indices. 

The allowed charge configurations determine Gauss Law constraints such that for each type of charge, 
$\exp(i A_l a_0)$ at the lattice scale creates the $l$th charge configuration with linear size $a_0$.
In particular, given a lattice discretization, it is trivial to write down a discrete version of Gauss' Law for 
which the action of $\exp(i A_l a_0)$ creates the $l$th charge configuration, and Taylor expansion 
leads to a continuum Gauss Law, of the general form
\begin{equation}
\sum_l D_l^a E_l(\bv{r}) = \rho_a(\bv{r})
\label{eqn:Gauss}
\end{equation}
where $D_l^a$ are linear differential operators, possibly containing dimensionful coupling constants reflecting UV physics.

In general, the geometry of the charge configurations determines the mobility of charges. For example, charges are 
mobile in the $x$ direction if and only if an $x$-oriented dipole is an allowed local charge configuration. We can thus conclude that
adding more charge configurations to the generating set can never decrease the mobility of charges. Furthermore, 
if the allowed charge configurations have only fractal structure, as in the $U(1)$ Haah's code model, then we generally expect that charges appear only at the cutoff scale.

Gauss' Law leads to a gauge transformation rule $A_l \rightarrow A_l - \tilde{D}_l^a \alpha^a$, and $\theta_a \rightarrow \theta_a - \alpha^a$, where $\theta_a$ is a phase variable
conjugate to the charge density $\rho_a$. 
$\tilde{D}_l^a$ is defined from $D_l^a$ by multiplying every term in $D_l^a$ that has $n$ derivatives by $(-1)^n$. 

Given the gauge transformation rules, we can define the ``magnetic field" of the theory. The $k$th component 
of the magnetic field takes the general form
\begin{equation}
B^k = \sum_l C^k_l A_l
\end{equation}
where the $C^k_l$ are differential operators. Gauge invariance requires that for every $a$,
\begin{equation}
\sum_l C^k_l \tilde{D}_l^a = 0
\label{eqn:BConstraint}
\end{equation}
as an operator equation. Every independent solution of these equations defines a component of the magnetic field. 
The index $k$ need not have anything to do with spacetime indices.

It is important to ask when solutions to Eq. \eqref{eqn:BConstraint} exist, because if there is no magnetic field 
then the theory becomes macroscopically degenerate. If there are $M$ Gauss laws and $N$ components of the electric field, we conjecture 
that $N>M$ is sufficient, except in certain degenerate cases, and ``generically" necessary for Eq. \eqref{eqn:BConstraint} 
to have a solution. A $\mathbb{Z}_p$ lattice version of this conjecture is stated precisely and proven in Ref. \cite{HaahModules}, and a $U(1)$ version is discussed in Ref. \cite{HaahU1Code}. We define our language more precisely and prove the conjecture for $M=1$ and $2$ in Appendix \ref{app:magneticFieldProof}.

After defining magnetic fields, it becomes possible to write down a free field gauge theory.
The corresponding Lagrangian density, to lowest order, is of the form
\begin{align}
\mathcal{L} = \frac{1}{2}\sum_i &\left(\sum_a \tilde{D}_i^a A_0^a + \partial_t A_i\right)^2 - \frac{1}{2}\sum_k (B^k)^2 + \nonumber \\
&+ \sum_{ik}\theta_{ik}(\sum_a \tilde{D}_i^a A_0^a+\partial_t A_i)B_k
\end{align}
where we have now allowed for the presence of a ``theta term." The theta term modifies Gauss' Law and, unlike in standard one-form gauge theory, 
can in general affect the equations of motion. 

As in the field theory corresponding to Haah's code, there can be many conserved quantities. 
In general, a globally conserved charge
\begin{align}
Q_{\lbrace f \rbrace} = \sum_a \int f^a(\bv{r})\rho_a(\bv{r})
\end{align}
exists, associated to any set of functions $\{f^a(\bv{r})\}$ which satisfy 
$\sum_a \tilde{D}^a_i f^a = 0$ for each $i$. 

In the ``noncompact'' case, where we take $A_l({\bf r})$ to be a real scalar field at the lattice scale, the theory is stable as long
as the photon dispersion is not flat. Such a model may arise through duality from a lattice system with $U(1)$ subsystem symmetry,
generalizing the familiar case of $U(1)$ particle-vortex duality in $(2+1)D$ gauge theory. However if we take $A_i$ to be a rotor
variable on the lattice scale (where $A_i \sim A_i + 2\pi$), then there are instanton processes, and the number of relevant 
instanton operators determines whether the phase is fully stable or corresponds to a (multi-)critical point. 
We leave a general study of such instanton processes to future work. 

\textit{Sierpinski Prism Model---} As a further example of this logic, we now also define a $U(1)$ gauge theory 
description of Yoshida's Sierpinski prism model \cite{YoshidaFractal}. An interesting property of this model is 
that it does not possess charge conservation; nevertheless, a generalized gauge theory can be used to describe the phase. 

\begin{figure}
	\includegraphics[width=8cm]{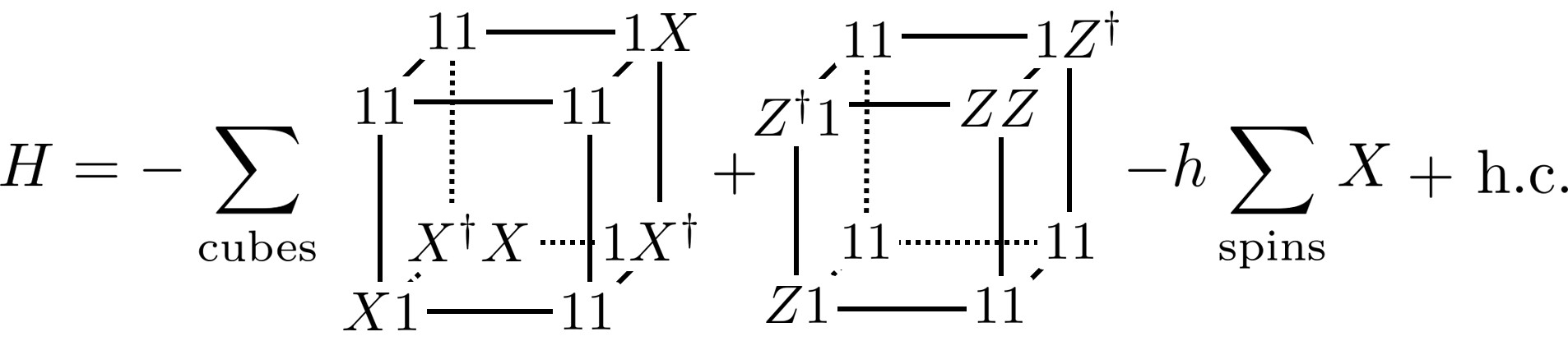}
	\caption{Hamiltonian for our chosen $\mathbb{Z}_N$ generalization of Yoshida's Sierpinski prism model. There is one of each term per elementary cube of the cubic lattice, two $N$-component spins per site, and each term is a product of five generalized Pauli operators. }
	\label{fig:YoshidaStabilizers}
\end{figure}

\begin{figure}
\centering
\hfill
\subfloat[\label{fig:HopCharges}]{\includegraphics[width=2cm]{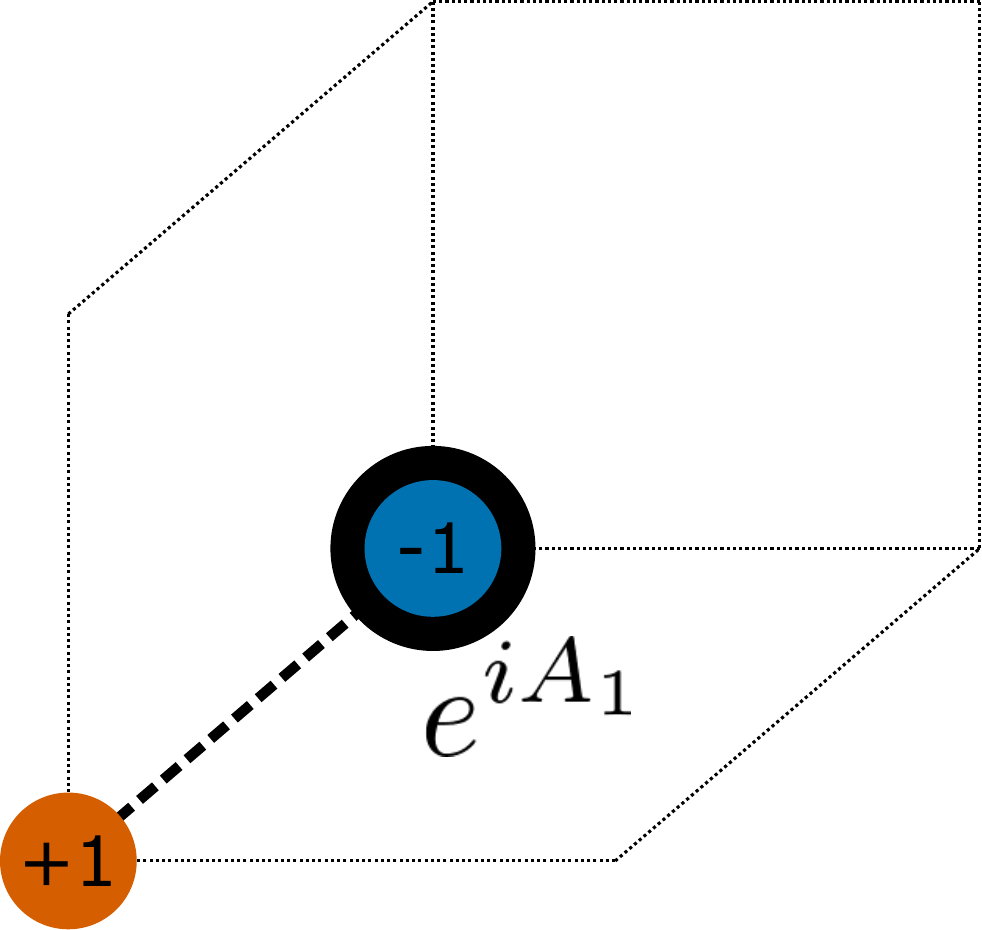}} \hfill
\subfloat[\label{fig:TriangleCharges}]{\includegraphics[width=2cm]{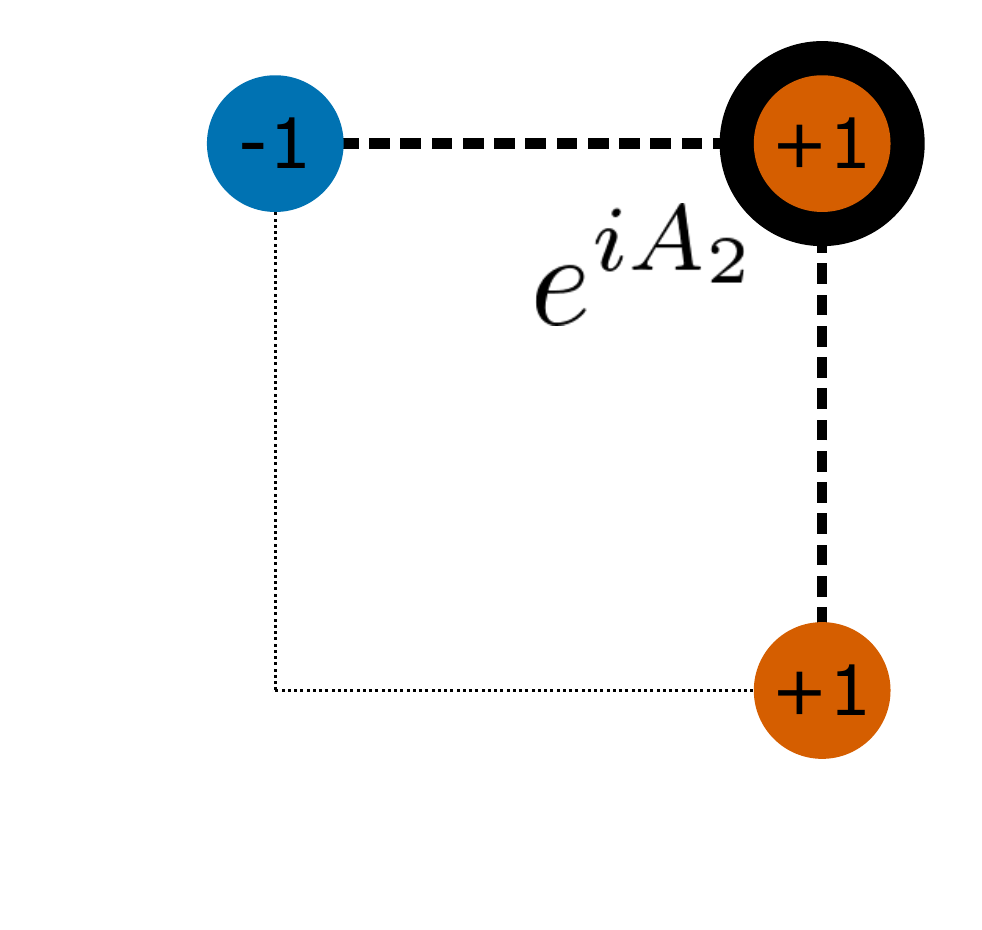}} \hfill
\subfloat[\label{fig:IsolatedYoshidaCharge}]{\includegraphics[width=4cm]{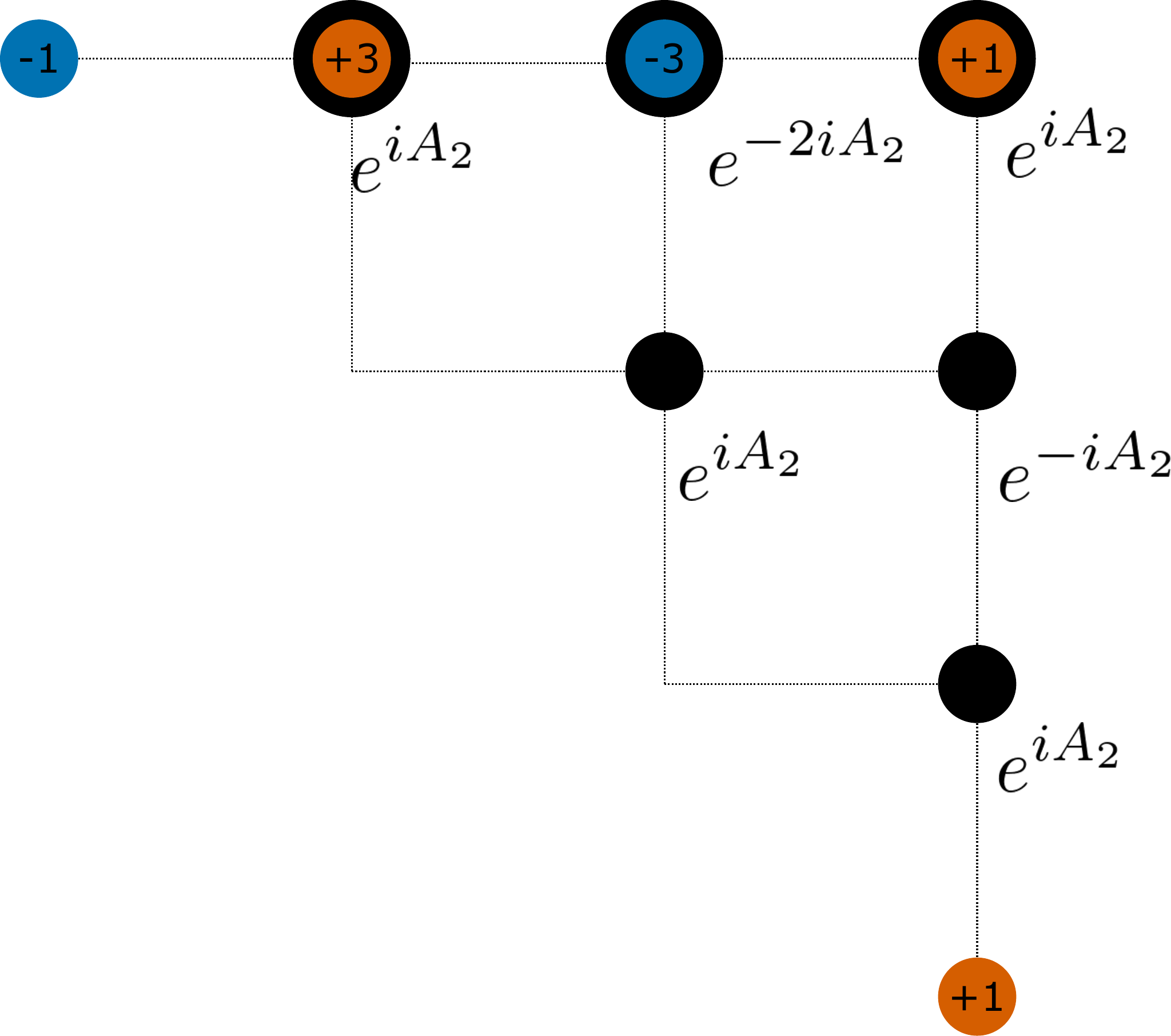}}
\caption{Charge configurations created in our $U(1)$ Sierpinski prism gauge theory. Color coding is the same as Fig. \ref{fig:HaahCharges}. (a) and (b): Generating charge configurations, created by the local action of $e^{iA_1}$ and $e^{iA_2}$ respectively. (c): Typical charge configuration with an isolated $+1$ charge created with repeated application of $e^{iA_2}$ (we have set $a_0=1$ for legibility). Isolating one charge in the $xy$-plane requires a line of charges of length $d$ to be created at distance $d$.}
\label{fig:YoshidaCharges}
\end{figure}

One $\mathbb{Z}_p$ generalization of the Sierpinski prism model, which has two $\mathbb{Z}_p$ spins on each site of 
the cubic lattice, has the Hamiltonian shown in Fig. \ref{fig:YoshidaStabilizers}. The generating set of charge configurations 
is shown in Figs. \ref{fig:HopCharges} and \ref{fig:TriangleCharges}. 
The generating charge configurations correspond to the continuum Gauss law
$D_1 E_1 + D_2 E_2 = \rho$, with $D_1 = \partial_z$,  $D_2 = a_0 \partial_x \partial_y + \partial_y + a_0^{-1}$,
where $a_0$, as before, has dimensions of length.

In this theory, Wilson lines $\exp(i\int dz A_1)$ create charges at their ends, but the in-plane ($xy$-plane) dynamics are fractal, similar to the 
$U(1)$ Haah's code theory. In-plane charge configurations are only created using operators at the cutoff scale because charges 
can only be isolated in-plane in the presence of additional lines of charges, as in Fig. \ref{fig:IsolatedYoshidaCharge}.

The appearance of a term $E_2$ with no derivatives in Gauss' Law is the manifestation of the lack of a globally 
conserved $U(1)$ charge. The magnetic field is
$B = \tilde{D}_2 A_1 - \tilde{D}_1 A_2$
and the photon dispersion is
$\omega^2 = a_0^{-2} + \left(k_y^2 -2k_xk_y + k_z^2\right) + a_0^2 k_x^2 k_y^2$.
Curiously, at $k_z=0$ and fixed $k_x$, the minimum gap occurs when $k_y = k_x/(1+a_0k_x^2)$, and its value is
$\Delta_{min}(k_x)=\frac{1}{a_0^2}\frac{1}{1+a_0^2 k_x^2}$,
which goes to zero at $k_x \rightarrow \infty$. Therefore the low-energy dynamics does not occur at long wavelengths, so even the gauge sector
of the field theory requires a lattice regularization. A non-zero $a_0$ means that there is a UV regulator and thus a momentum cutoff, so the 
photon will remain fully gapped.
A Higgsed version of this theory may be constructed in the same way as for Haah's code, yielding a field theoretic formulation of the 
gapped $\mathbb{Z}_p$ Sierpinski prism model.

\begin{figure}
\includegraphics[width=5cm]{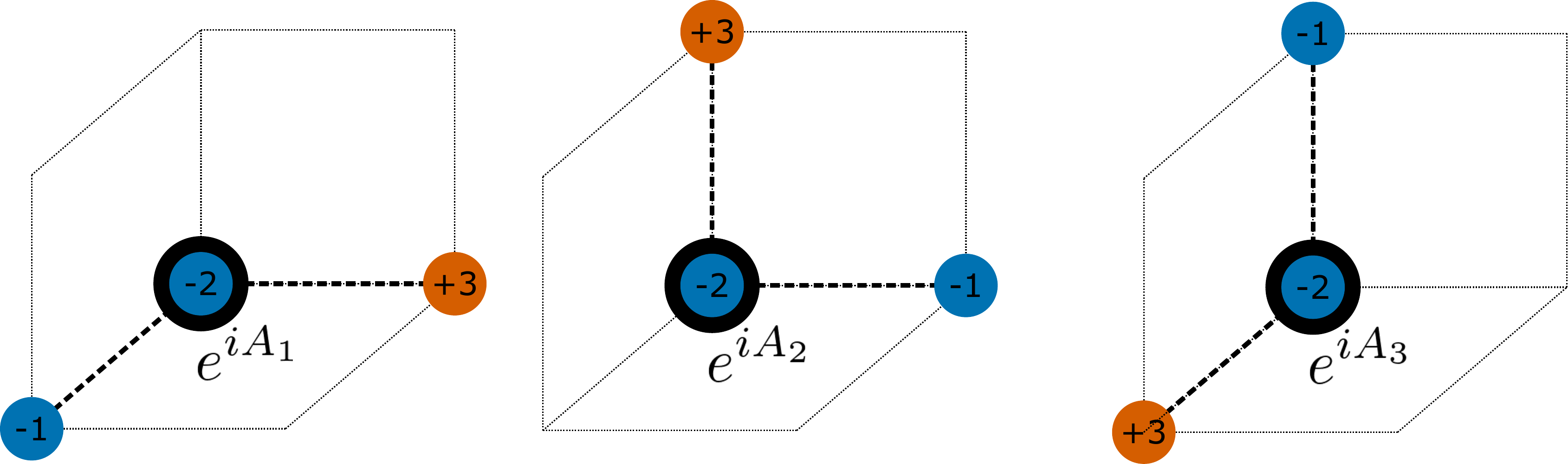}
\caption{Charge configurations created by local operators in our simple $U(1)$ model with Gauss law in Eq. \eqref{eqn:NewU1Gauss}.}
\label{fig:NewU1Charges}
\end{figure}

\textit{$U(1)$ model without $\mathbb{Z}_p$ counterpart--}
Our final example is a simple $(3+1)$D $U(1)$ model where all point-like excitations have fractal dynamics, but which has fully 
mobile excitations upon breaking the $U(1)$ gauge symmetry down to any discrete subgroup. This demonstrates 
that fractal dynamics at the $U(1)$ level need not be related to gapped fracton models, even in $(3+1)D$ (in $(2+1)D$ all lattice $\mathbb{Z}_p$ charges are mobile for $p$ prime \cite{HaahModules}). Our model has one charge 
type and three charge configurations, shown in Fig. \ref{fig:NewU1Charges}, leading to a Gauss law
\begin{equation}
\rho = (3\partial_x - \partial_z)E_1 + (3\partial_y - \partial_x)E_2 + (3\partial_z - \partial_y)E_3 
\label{eqn:NewU1Gauss}
\end{equation}

Inspection of the allowed charge configurations makes the fractal dynamics clear, as well as the mobility of the charges when 
charge-$p$ excitations are condensed (the Higgs phase does not have subdimensional particles). With our usual definitions for 
$D_i$, there are three magnetic field components $B_i = \epsilon_{ijk}\tilde{D}_jA_k$, with $\epsilon$ the Levi-Civita symbol, 
which obey the constraint $\tilde{D}_iB_i = 0$. The photon dispersion is $\omega^2 = 4k^2+3\sum_{i<j}(k_i-k_j)^2$, and 
the only conserved quantity is the total charge.

In this model it is particularly clear how the matter sector depends crucially on the lattice
regularization. Our discussion, as well as the charge configurations in Fig. \ref{fig:NewU1Charges}, assumed 
regularization on a cubic lattice. As shown in Appendix \ref{app:regularization}, if we 
were to regularize not on a cubic lattice but on a rhombohedral lattice with primitive lattice vectors $(3,0,-1)$, $(-1,3,0)$, 
and $(0,-1,3)$, the charge configurations would simply be dipoles along the primitive vectors, and we obtain standard 
$U(1)$ lattice gauge theory on a rhombohedral lattice. 

\textit{Acknowledgments---}We thank Jeongwan Haah for helpful comments on a draft. DB is supported by the Laboratory for Physical Sciences and Microsoft. MB is supported by 
NSF CAREER (DMR-1753240) and JQI-PFC-UMD. 

\appendix

\section{Proof of Magnetic Field Existence}
\label{app:magneticFieldProof}

In this appendix we prove that for $M=1$ or $2$ charge types, $N > M$ non-degenerate charge configurations is sufficient to have a well-defined, gauge-invariant magnetic field and discuss the situations when $N>M$ is necessary.

We label different charge types by $\lambda=\alpha,\beta,...$, label components of the magnetic field by $i,j,k... \in \lbrace 1, 2, ... N\rbrace $, and label spatial directions by $a,b,c...\in \lbrace x,y,z \rbrace$.

We first define ``non-degenerate." The first ``degenerate" case is when two field components create the same charge configuration (up to scalar multiples), that is, $D_i^{\lambda} = c D_j^{\lambda}$ for some nonzero scalar $c$, all $\lambda$, and some $i \neq j$. If this is the case, then $A_i - \lambda A_j$ is gauge-invariant and therefore should not be considered part of the gauge structure. More generally, in the space of charge configurations, a nondegenerate configuration requires all the generating configurations to be linearly independent.

The second degenerate case is when, for $M=2$, the two Gauss laws decouple. Mathematically, this occurs if, for every $i$, either $D_i^{\alpha}$ or $D_i^{\beta}$ is zero. Then obviously we must apply the $M=1$ result to each Gauss law separately. This situation can occur for larger $M$, but it is more subtle to make the condition mathematically precise. 

We will also make use of the fact that the ring of linear differential operators over the integers is an integral domain (no zero divisors) and a unique factorization domain. This can be seen via isomorphism to the ring of polynomials with integer coefficients.

Recall that a gauge-invariant magnetic field
\begin{equation}
B = \sum_{i=1}^N C_i A_i
\end{equation}
exists if and only if there is a solution to the set of $M$ equations
\begin{equation}
\sum_{i=1}^N \tilde{D}_i^{\lambda}C_i = 0
\label{eqn:gaugeInvCondition}
\end{equation}

We first discuss what happens when $N \leq M$. For $N=1$ the nonexistence of zero divisors means that no nonzero solution to $\tilde{D}_1^{\lambda}C_1=0$ exists, so magnetic fields can never be defined. For $1 < N \leq M$ a magnetic field can be defined in certain non-generic cases. Suppose, for example, that
\begin{equation}
\rho^{\lambda} = F^{\lambda}\sum_{i=1}^N G_i E_i
\label{eqn:specialCase}
\end{equation}
where $F^{\lambda}$ and $G_i$ are nonzero differential operators. Then, using tildes as in the main text, it is straightforward to show that $\tilde{G}_i A_j - \tilde{G}_j A_i$ is gauge-invariant for $i \neq j$. We claim that for $M=2$, this is the only condition under which $N \leq M$ permits a solution, but it is unclear if there are more such cases for $M>2$. 

To prove the claim, suppose $N=2$ and that there exists a nontrivial magnetic field. By similar arguments for the $N=1$ case, we must have $C_1,C_2 \neq 0$ and the $\tilde{D}_i^{\lambda}$ all nonzero. Applying $\tilde{D}_2^{\alpha}$ to the $\lambda=\beta$ version of Eq. \eqref{eqn:gaugeInvCondition} and $\tilde{D}_2^{\beta}$ to the $\lambda=\alpha$ version and subtracting, we obtain
\begin{equation}
C_1 \(\tilde{D}_1^{\beta} \tilde{D}_2^{\alpha}-\tilde{D}_1^{\alpha} \tilde{D}_2^{\beta}\) = 0
\label{eqn:assumedSol}
\end{equation}
By uniqueness of factorization, there exist linear differential operators $\tilde{F}_i$ and $\tilde{G}_i^{\lambda}$ such that $\tilde{D}_i^{\lambda} = \tilde{F}_i \tilde{G}_i^{\lambda}$ and $\tilde{G}_i^{\alpha}$ and $\tilde{G}_i^{\beta}$ are relatively prime. Plugging this into Eq. \eqref{eqn:assumedSol} and using both the lack of zero divisors and uniquness of factorization, we find $\tilde{G}_i^{\alpha} = \tilde{G}_i^{\beta} \equiv \tilde{G}_i$. That is, if a magnetic field exists for $M=N=2$, then Eq. \eqref{eqn:specialCase} is satisfied, as desired.

We now prove that $N>M$ is sufficient for a solution to exist for $M=1$ and $M=2$.

For $M=1$, we simply note that all the $\tilde{D}_i$ are nonzero by nondegeneracy. Hence, for any pair $i$ and $j$, we may take $C_i = \tilde{D}_j$, $C_j = -\tilde{D}_i$, and all other $C_k = 0$ to obtain a nontrivial solution.

We now exhibit a solution for $M=2$, $N=3$. (For $N>3$, one can simply ignore $A_4,A_5,...$ and apply the $N=3$ result). We prove this by exhibiting an ansatz for a magnetic field and showing that it fails only when Eq. \eqref{eqn:specialCase} holds, in which case we already know how to define a magnetic field.

By nondegeneracy, without loss of generality we may assume that $\tilde{D}_1^{\lambda}$ are both nonzero. We then make the ansatz
\begin{align}
C_1 &= \tilde{D}_1^{\alpha}\(\tilde{D}_2^{\beta} \tilde{D}_3^{\alpha}-\tilde{D}_2^{\alpha} \tilde{D}_3^{\beta}\) \nonumber\\
C_2 &= \tilde{D}_1^{\alpha}\(\tilde{D}_3^{\beta} \tilde{D}_1^{\alpha}-\tilde{D}_3^{\alpha} \tilde{D}_1^{\beta}\) \nonumber\\
C_3 &= \tilde{D}_1^{\alpha}\(\tilde{D}_1^{\beta} \tilde{D}_2^{\alpha}-\tilde{D}_1^{\alpha} \tilde{D}_2^{\beta}\)
\end{align}
It can be checked explicitly that this satisfies Eq. \eqref{eqn:gaugeInvCondition}. This ansatz fails to give a meaningful magnetic field only when these $C_i$ are all zero. Suppose that this is the case. Following similar arguments to the $M=2,N=2$ case, we can always decompose $\tilde{D}_i^{\lambda} = \tilde{F}_i \tilde{G}_i^{\lambda}$ where $\tilde{G}_i^{\alpha}$ and $\tilde{G}_i^{\beta}$ are relatively prime. Again using the uniqueness of factorization, we find that $C_i$ all zero forces $\tilde{G}_i^{\lambda} \equiv G^{\lambda}$ to be independent of $i$, which means that Eq. \eqref{eqn:specialCase} is satisfied.

\section{Regularization}
\label{app:regularization}

In this appendix we briefly explain the distinct regularization procedures for Eq. \eqref{eqn:NewU1Gauss}.

Consider discretizing Eq. \eqref{eqn:NewU1Gauss} on a cubic lattice of lattice constant $a_0$. Using forward derivatives, we find
\begin{align}
\rho(\bv{r})/a_0 = \sum_i &\left[3E_i(\bv{r}+ a_0\hat{\bv{x}}_i)-2E_i(\bv{r})\right] -E_1(\bv{r}+a_0\hat{\bv{z}}) \nonumber \\
&- E_2(\bv{r}+a_0\hat{\bv{x}}) - E_3(\bv{r}+a_0\hat{\bv{y}})
\end{align}
which, by inspection, leads to the charge configurations shown in Fig. \ref{fig:NewU1Charges}. On the other hand, we can transform to coordinates $u=3x-z$, $v=3y-x$, and $w=3z-y$, which form a rhombohedral coordinate system. In these coordinates (since $E_i$ are assumed not to transform), we obtain the continuum Gauss law
\begin{equation}
\rho = \partial_u E_1 + \partial_v E_2 + \partial_w E_3
\end{equation}
Lattice regularizing in these rhombohedral coordinates, we obtain
\begin{align}
\rho(\bv{r})/a_0' = &E_1(\bv{r}+a_0'\hat{\bv{u}}) + E_2(\bv{r}+a_0'\hat{\bv{v}})+\nonumber \\
& + E_3(\bv{r}+a_0'\hat{\bv{w}}) -\sum_i E_i(\bv{r})
\end{align}
where $a_0' = \sqrt{10}a_0$ accounts for the longer primitive lattice vectors. This Gauss law describes standard one-form $U(1)$ gauge theory on the rhombohedral lattice, albeit with the degrees of freedom located on the sites of the lattice.

\bibstyle{apsrev4-1} \bibliography{references}

\end{document}